\documentclass[journal=jpcbfk]{achemso} 
\usepackage{amsmath, amsfonts, amssymb, graphicx, subfigure, xcolor,indentfirst,lmodern,soul,url}
\usepackage[labelfont=bf]{caption} 

\author{Hao Tian}
\affiliation{Department of Chemistry, Center for Research Computing, Center for Drug Discovery, Design, and Delivery (CD4), Southern Methodist University, Dallas, Texas, United States of America}

\author{Sian Xiao}
\affiliation{Department of Chemistry, Center for Research Computing, Center for Drug Discovery, Design, and Delivery (CD4), Southern Methodist University, Dallas, Texas, United States of America}

\author{Xi Jiang}
\affiliation{Department of Statistics, Southern Methodist University, Dallas, Texas, United States of America}

\author{Peng Tao}
\affiliation{Department of Chemistry, Center for Research Computing, Center for Drug Discovery, Design, and Delivery (CD4), Southern Methodist University, Dallas, Texas, United States of America}
\email{ptao@smu.edu}

\title{PASSerRank: Prediction of Allosteric Sites with Learning to Rank}

\begin{document}

\begin{abstract}

Allostery plays a crucial role in regulating protein activity, making it a highly sought-after target in drug development. One of the major challenges in allosteric drug research is the identification of allosteric sites. In recent years, many computational models have been developed for accurate allosteric site prediction. Most of these models focus on designing a general rule that can be applied to pockets of proteins from various families. In this study, we present a new approach using the concept of Learning to Rank (LTR). The LTR model ranks pockets based on their relevance to allosteric sites, i.e., how well a pocket meets the characteristics of known allosteric sites. The model outperforms other common machine learning models with higher F1 score and Matthews correlation coefficient. After the training and validation on two datasets, the Allosteric Database (ASD) and CASBench, the LTR model was able to rank an allosteric pocket in the top 3 positions for 83.6\% and 80.5\% of test proteins, respectively. The trained model is available on the PASSer platform (\url{https://passer.smu.edu}) to aid in drug discovery research. 

\end{abstract}

\section{Introduction}

Allostery is a biological process where an effector molecule binds to an allosteric site that is distant to the active site of a protein. This binding results in conformational and dynamic changes that can regulate the protein's function, making it a key aspect of cellular signaling and is considered as the second secret of life. \cite{liu2016allostery, wodak2019allostery, krishnan2022probing, fenton2008allostery} Despite its importance, the allosteric mechanisms of most proteins remain elusive. A universal protein allosteric mechanism has yet to be forumated. \cite{nussinov2013allostery,xiao2022machine}

Allostery offers several advantages in drug development. Compared to orthosteric site binding, allosteric site binding provides a controlled regulation of protein function that can either activate or inhibit the binding of ligands at orthosteric sites. \cite{peracchi2011exploring} Additionally, allosteric modulators are reported to have fewer side effects with no additional pharmacological effects once allosteric sites are saturated. \cite{wu2022prediction} Furthermore, allosteric sites experience low evolutionary pressure, ensuring the safety of on-target drugs. \cite{christopoulos2004g, de2014allosteric} These benefits make allosteric drug development a promising field and offer substantial advantages over orthosteric drug development.

Identifying appropriate allosteric sites is a major challenge in allosteric drug development. \cite{lu2014recent, lu2019allosteric} In recent years, numerous computational methods for allosteric site identification and prediction have been developed. With the help of machine learning (ML), Allosite \cite{huang2013allosite} applies support vector machine (SVM) to learn the physical and chemical features of protein pockets. Another ML-based approach, the three-way random forest (RF) model developed by Chen \textit{et al.} \cite{chen2016random}, is capable of predicting allosteric, regular, or orthosteric sites. PASSer \cite{tian2021passer, xiao2021passer2,tian2023passer} is a recently developed method that combines extreme gradient boosting (XGBoost) \cite{chen2016xgboost} with a graph convolutional neural network \cite{kipf2016semi} to learn physical and topological properties without any prior information. In addition to ML, traditional methods such as normal mode analysis (NMA) \cite{panjkovich2012exploiting} and molecular dynamics (MD) \cite{laine2010use} are widely used to investigate the communication between regulatory and functional sites, including SPACER \cite{goncearenco2013spacer} and PARS \cite{panjkovich2014pars}.

It is important to note the development of allostery databases, including the Allosteric Database (ASD), \cite{huang2011asd} which contains 1949 entries of protein-modulator complexes with annotated allosteric residues, and ASBench, \cite{huang2015asbench} a smaller benchmark set optimized from the ASD data. CASBench is a benchmarking set that includes annotated catalytic and allosteric sites. \cite{zlobin2019casbench}  These datasets play a crucial role in training and evaluating allosteric site prediction models.

Most previous research on prediction models focused on developing universal models for allosteric site prediction. These models intend to make ``absolute'' predictions (either as labels or probabilities) for all pockets detected in different types of proteins, which is a challenging and time-consuming task. Learning to Rank (LTR), as an emerging area, was first applied in information retrieval \cite{trotman2005learning} and has been used in many bioinformatics studies, ranging from drug-target interaction prediction \cite{ru2022nerltr} to compound virtual screening \cite{furui2022compound}. Unlike ``absolute'' predictions, LTR models provide ``relative'' predictions by ranking objects from the most to the least relevant to a target, making it a more achievable and reasonable approach for allosteric site prediction.

In this study, we present the state-of-the-art machine learning model on allosteric site prediction with LTR. The LTR model is implemented using LambdaMART. LambdaMART combines gradient boosting decision tree (GBDT) with the loss function derived from LambdaRank, a LTR algorithm. Compared with other machine learning models such as XGBoost, SVM, and RF, LambdaMART achieved the highest F1 score and Matthews correlation coefficient (MCC). Moreover, this model has a better ability to rank actual allosteric sites at top positions. The trained LambdaMART model is freely available at PASSer (\url{https://passer.smu.edu}) to facilitate related research.

\section{Methods}

\subsection{Allosteric Protein Databases}

Two databases were used to train and validate different machine learning models, including the Allosteric Database (ASD) and CASBench. 

In the latest version of ASD, there are 1949 entries of protein-modulator complexes. To ensure data quality, a clearning process is applied to the protein-modulator complexes based on standards proposed in the Allosite study \cite{huang2013allosite}. Three standards, including high-resolution protein structures with a resolution smaller than 3 $\mathring{A}$, the presence of a complete structure in the allosteric site, and a low sequence identity threshold of 30\%, was applied to select high-quality and sequence-diverse proteins in the overall training set. If two or more proteins have high sequence identity, the one with the shortest modulator-pocket distance is retained to ensure the finest labeling. The modulator-pocket distance calculation is described below in Section \ref{pocket}. A total of 207 proteins were selected in the overall training set and were randomly split into a training set (80\%) and a test set(20\%). To facilitate the cleaning process, a data processing pipeline script has been created and made available as open source on GitHub (\url{https://github.com/smu-tao-group/PASSerRank}). 

The CASBench dataset was used as an external test set. The CASBench benchmark set comprises proteins annotated with allosteric sites, but only those entries that include both allosteric ligands and sites were included. Additionally, proteins that were already present in the ASD dataset were removed to ensure the validity of the benchmark set. 

\subsection{Pocket Descriptors and Labeling}
\label{pocket}

FPocket is an open-source software for protein pocket detection. In this work, FPocket was applied on each protein to detect protein pockets. On average, 21 pockets were detected in each protein, with a total of 4413 pockets in 207 proteins. For each detected pocket, 19 physical and chemical features are calculated, ranging from pocket volume, solvent accessible surface area to hydrophobicity. A complete list of feature names is shown in Figure \ref{fig:shap}.

To label each pocket as an allosteric or non-allosteric site, we have automated the process by assigning the closest pockets to the modulator as the allosteric site. The center of mass is first calculated for all pockets and the modulator, and then the pairwise distances between the pockets and the modulator are computed. The pocket with the shortest distance is labeled as positive (allosteric site), while all other pockets are labeled as negative (non-allosteric site). However, if the closest distance is greater than 10 $\mathring{A}$, this entry is removed from the dataset, as such a large distance may indicate inaccurate pocket detection and negatively impact model performance.

\subsection{Learning to Rank}

Prior researches on allosteric site prediction focus on developing a universal model that can accurately predict allosteric sites in all proteins. However, in practice, it is more important to identify the most promising pockets within each individual protein. Therefore, a machine learning model that is capable of ranking pockets in order of their likelihood to be allosteric sites is more desirable and attainable than a binary classification model that provides absolute predictions for all pockets.

In this study, we implemented the LTR algorithm using GBDT and the LambdaMART method. GBDT is a popular machine learning approach that iteratively learns from decision trees and ensembles of their predictions. Here, we use LightGBM \cite{ke2017lightgbm}, one of the two popular implementations of GBDT, over XGBoost \cite{chen2016xgboost}. LambdaMART is an LTR method that trains GBDT with the lambdarank loss function. The lambdarank loss function optimizes the value of the normalized discounted cumulative gain (NDCG) for the top $K$ cases, and is calculated using discounted cumulative gain (DCG) and ideal discounted cumulative gain (IDCG) as:
\begin{align}
\text{DCG}@K &= \sum_{i=1}^K \frac{2^{G_i} - 1}{\log_2 (i + 1)}\\
\text{IDCG}@K &= \sum_{i=1}^K \frac{2^{|G|_i} - 1}{\log_2 (i + 1)}\\
\text{NDCG}@K &= \frac{\text{DCG}@K}{\text{IDCG}@K}
\end{align}
where $G_i$ is the gain (graded relevance value) at position $i$ and $|G|$ is the ideal ranking.

The LGBMRanker module in the LightGBM package (v3.3.4) was used to implementate the LambdaMART algorithm with GBDT as boosting type and lambdarank as the objective function.

\subsection{Machine Learning Models}

In addition to the LTR model, other commonly used machine learning models in allosteric site prediction were considered for comparison. XGBoost and RF are tree-based models. As previously stated, XGBoost is an implementation of the GBDT model that could also be used to train the LTR model. The RF model employs a bagging approach, training several independent decision trees in parallel. The prediction of RF is obtained through the weighted average of the outputs of all decision trees. The SVM classifier, on the other hand, learns a high-dimensional hyperplane that separates data points based on their labels. The XGBoost algorithm was implemented using the XGBoost package (version 1.7.3), and the RF and SVM classifiers were implemented using the Scikit-learn package (version 1.2.0). \cite{pedregosa2011scikit}

SHapley Additive exPlanations (SHAP) value is a method to increase model interpretability by quantifying feature importance. It has been implemented recently to explain tree-based models. \cite{yin2023unveiling} In this study, the SHAP values of 19 features from FPocket were calculated and compared. The method is implemented in the SHAP package (v0.41.0). \cite{lundberg2020local} 

\subsection{Performance Criteria}

Several metrics were calculated to compare and evaluate different machine learning models. Precision, recall, and specificity are good indicators for binary classification. The F1 score is a weighted measure of precision and recall. Moreover, it is reported that the Matthews correlation coefficient is a more suitable indicator than the F1 score and accuracy in binary classification evaluation \cite{chicco2020advantages}. 

\begin{align}
\text{Precision} &= \text{TP} \: / \: (\text{TP} \: + \: \text{FP}) \label{eq:precision} \\
\text{Recall} &= \text{TP} \: / \: (\text{TP} \: + \: \text{FN}) \label{eq:recall} \\
\text{Specificity} &= \text{TN} \: / \: (\text{TN} \: + \: \text{FP}) \\
\text{F1 score} &= 2 * \text{Precision} * \text{Recall} \: / \: (\text{Precision} \: + \: \text{Recall}) \\
\text{Accuracy} &= (\text{TP} \: + \: \text{TN}) \: / \: (\text{TP} \: + \: \text{FP} \: + \: \text{FN} \: + \: \text{TN}) \\
\text{MCC} &= \frac{\text{TP} * \text{TN} \: - \: \text{FP} * \text{FN}}{\sqrt{(\text{TP} \: + \: \text{FP}) * (\text{TP} \: + \: \text{FN}) * (\text{TN} \: + \: \text{FP}) * (\text{TN} \: + \: \text{FN})}}
\end{align}

The percentage of actual allosteric sites that are ranked in the top 1, 2, and 3 positions is calculated. This metric is commonly used in evaluating various allosteric site prediction models. The actual allosteric sites are compared with the predicted top 3 most probable pockets in each protein, and the percentage is calculated and accumulated for each position.

\section{Results}

\begin{figure}[t]
\centering
\includegraphics[scale=3.]{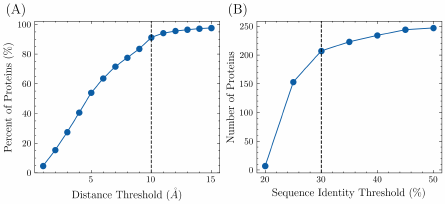}
\caption{The number of proteins included in the training set, along with different distance and sequence identity thresholds. (A) The minimum distances between the center of masses from pockets to the modulator in each protein were calculated. A protein-modulator complex is discarded if the minimum distance is higher than the threshold. With the threshold being set as 10$\mathring{A}$, 91.1\% of proteins were included. (B) To ensure uniqueness, proteins with high sequence identity were removed. The threshold was set as 30\%. A total of 207 proteins are included in the training set.}
\label{fig:identity}
\end{figure}

In this study, we chose three established standards and the pocket labeling strategy to prepare the training data for machine learning models. To ensure the quality of the protein-modulator complexes, we only considered those with high resolution protein structures (i.e., with a resolution of less than 3$\mathring{A}$) as reported in the RCSB Protein Data Bank. \cite{burley2017protein} Any protein structures with missing modulators were excluded from the analysis.

FPocket was used to identify potential pockets in each protein. The center of mass was calculated for both the identified pockets and the modulator. The pairwise distances between the center of mass of each pocket and the modulator were calculated. The closest pocket to the modulator was designated as the allosteric site while the other pockets were designated as non-allosteric sites. To ensure that the allosteric site and modulator are in contact, a distance threshold was imposed on the closest pocket. The effect of different distance thresholds on the percentage of proteins included in the training set is shown in Figure \ref{fig:identity}(A), with a final distance threshold of 10 $\mathring{A}$ chosen to avoid the inclusion of incorrectly labeled pockets. Consequently, 91.1\% of proteins from the ASD were included in the training set. To avoid over-representation of highly similar proteins, the pairwise sequence similarity was calculated between each newly selected protein and all previously selected proteins. If the similarity was higher than a specified threshold, the protein structure was discarded. The effect of different sequence identity thresholds is shown in Figure \ref{fig:identity}(B), with a final threshold of 30\% chosen. After these steps, 207 proteins were included in the overall training set.

We randomly selected 80\% of these proteins as the training set and used the remaining 20\% as the testing set. A total of four machine learning models, including LambdaMART, XGBoost, random forest, and SVM, were trained through 5-fold grid search with cross validation. The grid search takes an exhaustive search strategy over all combinations of pre-specified parameter values. All models were trained on a high-performance-computing platform with a 60GB V100 graphical processing unit (GPU). The parameters were fine-tuned and determined with the best performance on the training set. For comparison, the performance of FPocket is reported, in which the pocket with the highest score was treated as the positive (allosteric) prediction and others as negative (non-allosteric) predictions based on FPocket results. Similarly for the LambdaMART predictions, the pocket with the highest prediction score in each protein was labeled as positive. This explains that precision and recall metrics have the same number in LambdaMART and FPocket models, respectively, as there is only one positive prediction. If this positive prediction is wrong, we have a false positive, and there will also be a false negative, leading to the same number of FP and FN and thus the same value of precision and recall defined in Equation \ref{eq:precision} and \ref{eq:recall}. 

\begin{table}[t]
\centering
\caption{Performance comparison of machine learning models on the ASD dataset.}
\label{tab:asd}
\begin{tabular}{cccccc}
\hline
Metric & LambdaMART & XGBoost & RF & SVM & FPocket\\
\hline
Precision & \textbf{0.662} $\uparrow$ & 0.586 $\uparrow$ & 0.528 $\downarrow$ & 0.444 $\downarrow$ & 0.556\\
Accuracy & \textbf{0.968} $\uparrow$ & 0.961 $\uparrow$ & 0.956 $\downarrow$ & 0.944 $\downarrow$ & 0.958\\
Recall & 0.662 $\uparrow$ & 0.609 $\uparrow$ & 0.677 $\uparrow$ & \textbf{0.758} $\uparrow$ & 0.556\\
Specificity & \textbf{0.983} $\uparrow$ & 0.979 $\uparrow$ & 0.970 $\downarrow$ & 0.953 $\downarrow$ & 0.978\\
F1 score & \textbf{0.662} $\uparrow$ & 0.596 $\uparrow$ & 0.593 $\uparrow$ & 0.559 $\uparrow$ & 0.556\\
MCC & \textbf{0.645} $\uparrow$ & 0.577 $\uparrow$ & 0.575 $\uparrow$ & 0.554 $\uparrow$ & 0.536\\
\hline
Top 1 & \textbf{59.5\%} $\uparrow$ & 56.6\% $\uparrow$ & 58.0\% $\uparrow$ & 57.5\% $\uparrow$ & 55.6\% \\
Top 2 & \textbf{73.9\%} $\uparrow$ & 69.6\% $\downarrow$ & 71.0\% $\downarrow$ & 69.6\% $\downarrow$ & 71.5\% \\
Top 3 & \textbf{83.6\%} $\uparrow$ & 80.7\% $\uparrow$ & 79.7\% $\uparrow$ & 78.3\% $\uparrow$ & 76.8\% \\
\hline
\end{tabular}
\end{table}

All models were evaluated using the testing set of ASD. The results are listed in Table \ref{tab:asd}. The percentage of true allosteric sites that appeared in the predicted top 1, 2, and 3 positions was calculated and abbreviated as Top 1, 2, and 3, respectively. The performance of four machine learning models was compared with FPocket. Both LambdaMART and XGBoost exhibited better performance than FPocket under all or most metrics. RF and SVM were comparable to FPocket with higher F1 scores, MCC, and Top 3 percentage. LambdaMART achieved the best performance in 8 out of 9 metrics among all models.

\begin{table}[t]
\centering
\caption{Performance comparison of machine learning models on the CASBench dataset.}
\label{tab:cas}
\begin{tabular}{cccccc}
\hline
Metric & LambdaMART & XGBoost & RF & SVM & FPocket\\
\hline
Precision & \textbf{0.608} $\uparrow$ & 0.504 $\downarrow$ & 0.431 $\downarrow$ & 0.395 $\downarrow$ & 0.550 \\ 
Accuracy & \textbf{0.963} $\uparrow$ & 0.953 $\downarrow$ & 0.941 $\downarrow$ & 0.932 $\downarrow$ & 0.956 \\
Recall & 0.608 $\uparrow$ & 0.657 $\uparrow$ & 0.767 $\uparrow$ & \textbf{0.803} $\uparrow$ & 0.550 \\
Specificity & \textbf{0.980} $\uparrow$ & 0.968 $\downarrow$ & 0.950 $\downarrow$ & 0.939 $\downarrow$ & 0.977 \\
F1 score & \textbf{0.608} $\uparrow$ & 0.569 $\uparrow$ & 0.551 $\uparrow$ & 0.529  $\downarrow$& 0.550 \\
MCC & \textbf{0.589} $\uparrow$ & 0.551 $\uparrow$ & 0.548 $\uparrow$ & 0.534 $\uparrow$ & 0.527 \\
\hline
Top 1 & 56.3\% $\uparrow$ & 52.5\% $\downarrow$ & 44.1\% $\downarrow$ & \textbf{57.3\%} $\uparrow$ & 55.5\% \\
Top 2 & \textbf{73.7\%} $\uparrow$ & 70.0\% $\downarrow$ & 68.4\% $\downarrow$ & 73.2\% $\uparrow$ & 71.4\% \\
Top 3 & \textbf{80.5\%} $\uparrow$ & 77.0\% $\uparrow$ & 76.6\% $\downarrow$ & 76.0\% $\downarrow$ & 76.7\% \\
\hline
\end{tabular}
\end{table}

These models were further evaluated using the CASBench dataset. The CASBench training data set was prepared with the same procedures as the ASD training data. In addition, the proteins included in the ASD training data were excluded in the CASBench set to ensure the evaluation validity. The same metrics were calculated, and the results are listed in Table \ref{tab:cas}. Compared with the numbers reported in Table \ref{tab:asd}, the performance of all models was decreased but within an acceptable range. Overall, LambdaMART is superior to FPocket and leads in 7 out of 9 metrics. This demonstrates the ability of LambdaMART to rank protein pockets in terms of the relevance to allostery, which leads to a high F1 score, MCC, and Top 3 percentage.

\begin{figure}[t]
\centering
\includegraphics[scale=2.0]{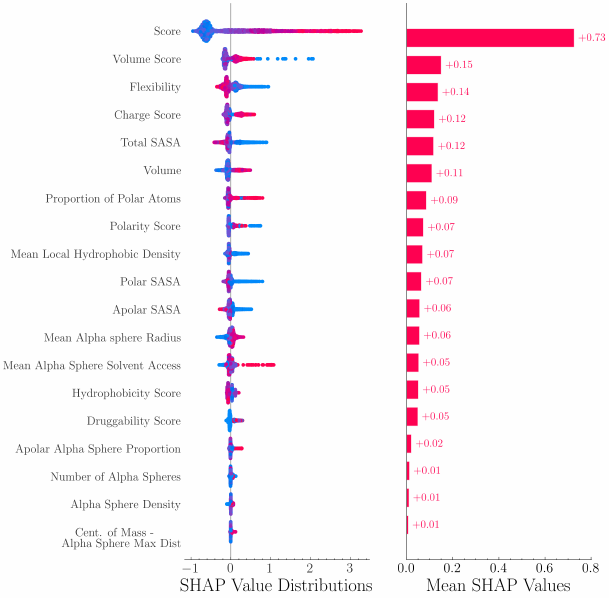}
\caption{SHAP value distributions and mean values of 19 features. These features are calculated from FPocket. Red and blue colors indicate positive and negative samples, respectively. FPocket score was identified as the most important feature.}
\label{fig:shap}
\end{figure}

The feature importance of the LambdaMART model was analyzed using SHAP values. As shown in Figure \ref{fig:shap}, the SHAP value distributions and mean SHAP values were displayed in descending order. Figure \ref{fig:shap} shows the distribution and mean SHAP values of the features in descending order. The results indicate that the FPocket score was the most important feature and significantly outperformed all other features. This highlights the effectiveness of the FPocket score in differentiating between allosteric and non-allosteric sites. Other features that were found to be important include the volume score, flexibility, charge score, and total solvent-accessible surface area (SASA). As seen from the SHAP value distribution, allosteric sites (represented in red) tend to have high FPocket scores, high volume scores, high charge scores, but low flexibility and low total SASA.

\begin{figure}[t]
\centering
\includegraphics[scale=0.5]{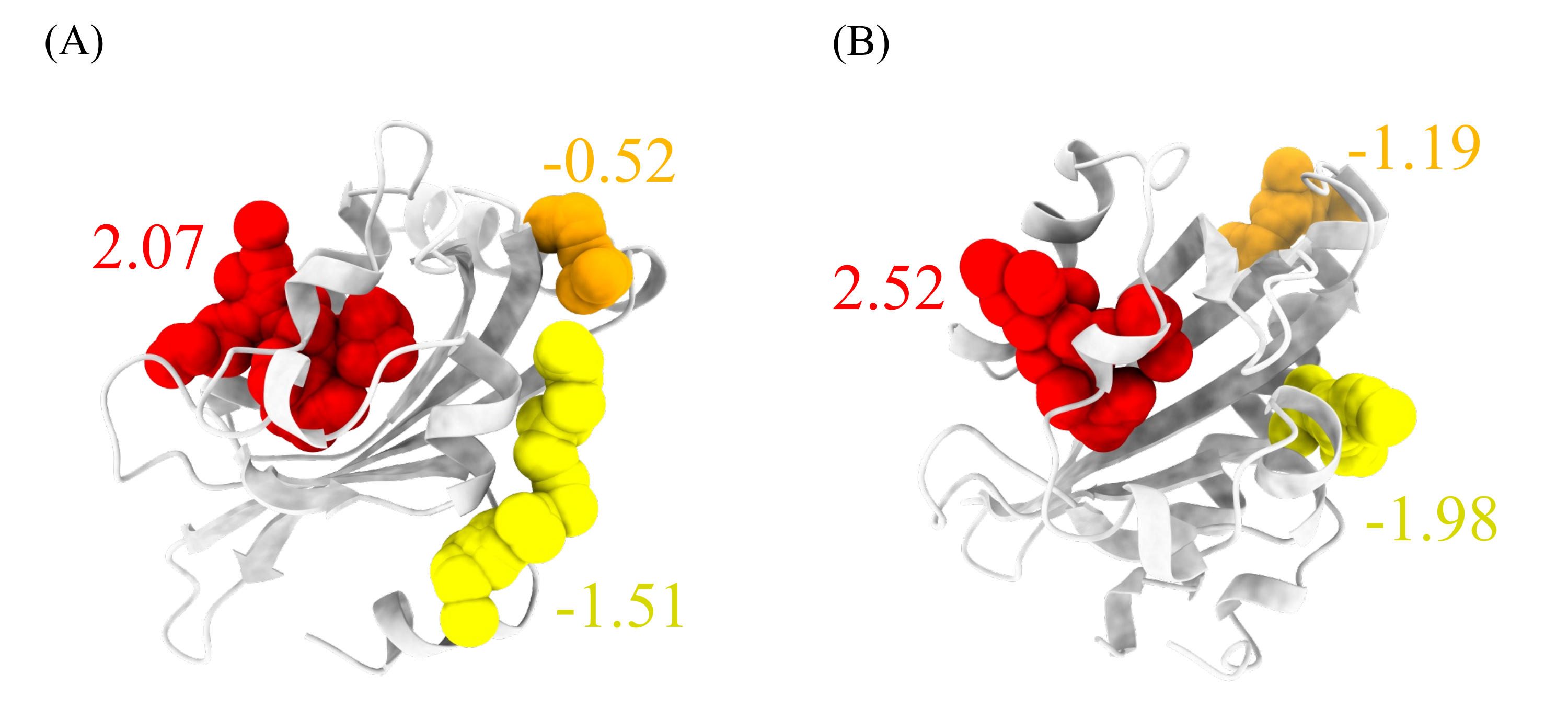}
\caption{Predictions of the light-oxygen-voltage domains from (A) \textit{Phaeodactylum tricornutum} Aureochrome 1a and (B) \textit{Avena Sativa} phototropin 1. The top 3 pockets are highlighted in red, orange, and yellow colors, respectivley. The top 1 pockets from both examples illustrated in red are the actual allosteric sites.}
\label{fig:example}
\end{figure}

The trained LambdaMART model has been made accessible to the public through the PASSer platform (\url{https://passer.smu.edu}). Users can access the model either through the webpage or through the command line interface using the PASSer API (\url{https://passer.smu.edu/apis/}). To demonstrate the efficacy of the model, two examples that were not part of the training and validation sets are presented in Figure \ref{fig:example}. These examples show the predicted allosteric sites of the light-oxygen-voltage domains of \textit{Phaeodactylum tricornutum} Aureochrome 1a \cite{tian2020deciphering} and \textit{Avena Sativa} phototropin 1 \cite{ibrahim2022dynamics} obtained using the LambdaMART model. The top three pockets are highlighted in red, orange, and yellow, respectivley, and with the corresponding predicted relevance scores. The top 1 predicted pockets illustrated in red are actual allosteric sites in both cases.

\section{Discussion}

The collection and cleaning of training data is a crucial step for the development of a high-performing machine learning model. The study by Huang \textit{et al.} \cite{huang2013allosite} applied three rules to select protein structures from the ASD dataset and curated a training set of 90 proteins. However, there is no available script to automate this process, which can result in an unfair model comparison. To address this issue, an open-source script to prepare machine learning-ready dataset is provided. This pipeline offers a simple and customizable benchmark preparation for evaluating various machine learning models. It should be noted that in the current design, proteins with multiple modulators in the same chain are discarded, as this can result in inconsistent ratios of allosteric and non-allosteric sites in each protein. Further refinement of the data cleaning process can lead to higher-quality training data.

Efforts have been invested in developing a universal model for allosteric site prediction by learning pockets from different proteins without considering the protein itself, such that all detected pockets from proteins in the training set are gathered and shuffled in a pool for training purposes. This approach, however, poses a challenge in model design and requires a model to learn a general rule that applies to all proteins of various families. Additionally, this training process is not reflective of real-world applications, where all pockets in a target protein need to be compared to determine the most probable ones. In light of these challenges, we offer a new perspective to rank pockets in each protein. The model focuses on the protein level and learns a ranking pattern among pockets. The proposed LambdaMART model outperforms other popular machine learning models such as XGBoost and SVM, with high F1 score and MCC, and is capable of ranking actual allosteric sites at the top positions. This demonstrates that it is more effective to learn the relative differences among pockets rather than a universal law applicable to all proteins.

In the context of allosteric site prediction, explainable machine learning is important as it helps researchers understand how a model arrives at its predictions. This information can be useful in drug design, as it can provide insights into the influencing factors that whether a pocket is likely to be an allosteric site. Tree-based models, such as random forest and gradient boosting decision tree, have good explainability as they can use metrics like Gini impurity to determine feature importance. SHAP values, a method from cooperative game theory, can also be used to quantify the contribution of each feature to the predictions made by a machine learning model. In this study, the SHAP values were used to indicate that the FPocket score was the most crucial feature, which aligns with the good performance of FPocket as a benchmark model. \cite{xiao2021passer2} The SHAP values also revealed that the model tends to predict pockets with high charge, volume, and low flexibility as allosteric sites, which can benefit the development of allosteric drugs.

\section{Conclusion}

The prediction of allosteric sites is crucial to the development of allosteric drugs. While many efforts have been dedicated to constructing a universal model for such prediction, this study presents a novel approach by employing a ranking model through the learning to rank concept. The proposed model outperforms other machine learning models based on various performance metrics, including a high rate of ranking true allosteric sites at top positions. Furthermore, a customizable pipeline is provided for the preparation of high-quality proteins for training purposes. The trained model is deployed on the PASSer platform (\url{https://passer.smu.edu}) and is readily available for public usage.


\section*{Data availability}
The authors declare that all data supporting the findings of this study are available within the paper.

\section*{Code availability}
The PASSer server is available at \url{https://passer.smu.edu}. The code to reproduce the training data and results is available at \url{https://github.com/smu-tao-group/PASSerRank}.

\section*{Competing interests}
The authors declare no competing interests.

\begin{acknowledgement}
Computational time was generously provided by Southern Methodist University's Center for Research Computing. Research reported in this paper was supported by the National Institute of General Medical Sciences of the National Institutes of Health under Award No. R15GM122013.

\end{acknowledgement}

\bibliography{cas-refs.bib}

\end{document}